\documentstyle[12pt]{article}

\begin{document}
\parskip 1ex
\setcounter{page}{1}
\oddsidemargin 0pt
\evensidemargin 0pt
\topmargin -40pt
%
\newcommand{\be}{\begin{equation}}
\newcommand{\ee}{\end{equation}}
\newcommand{\bea}{\begin{eqnarray}}
\newcommand{\eea}{\end{eqnarray}}
\def\a{\alpha}
\def\b{\beta}
\def\g{\gamma}
\def\G{\Gamma}
\def\d{\delta}
\def\e{\epsilon}
\def\z{\zeta}
\def\h{\eta}
\def\th{\theta}
\def\k{\kappa}
\def\l{\lambda}
\def\L{\Lambda}
\def\m{\mu}
\def\n{\nu}
\def\x{\xi}
\def\X{\Xi}
\def\p{\pi}
\def\P{\Pi}
\def\r{\rho}
\def\s{\sigma}
\def\S{\Sigma}
\def\t{\tau}
\def\f{\phi}
\def\F{\Phi}
\def\c{\chi}
\def\w{\omega}
\def\W{\Omega}
\def\de{\partial}

\def\pct#1{(see Fig. #1.)}

\begin{titlepage}
\hbox{\hskip 12cm NIKHEF/2003-001  \hfil}
\hbox{\hskip 12cm hep-th/0301021 \hfil}
\vskip 1.4cm
\begin{center}  {\Large  \bf  Truncations \ of \ the \  D9-Brane  \ Action  \vskip
.6cm and  \ Type-I  \ Strings}

\vspace{1.8cm}
 
{\large \large  Fabio Riccioni}
\vspace{0.6cm}

{\sl NIKHEF \\
Kruislaan 409 \ \ 1098 \ SJ \ \ Amsterdam \\
The Netherlands}
\end{center}
\vskip 1.5cm

\abstract{The low-energy effective action of type-I superstring theory 
in ten dimensions is obtained performing a truncation of type-IIB supergravity
in a background where D9-branes are present. 
The open sector corresponds to the first order in the low-energy 
expansion of the D9-brane action in a type-I
background. In hep-th/9901055 it was shown that there are two ways of performing 
a type-I truncation of the D9-brane action, and the resulting truncated action 
was obtained in a flat background. We extend this result to a generic type-I background,
and argue that the two different truncations are in correspondence with 
the open sector of the low-energy effective action of the two different consistent
ten-dimensional type-I string theories, namely the $SO(32)$ superstring and the 
$USp(32)$ non-supersymmetric string.}

\vskip 2.5cm
\begin{center}
{( January , \ 2003 )}
\end{center}
\vfill
\end{titlepage}
\makeatletter
\@addtoreset{equation}{section}
\makeatother
\renewcommand{\theequation}{\thesection.\arabic{equation}}
\addtolength{\baselineskip}{0.3\baselineskip} 

\section{Introduction}
Type-I string theory is obtained from type-IIB through an 
orientifold projection \cite{augusto} that removes the states that
are odd under orientation reversal of the string. From a 
target space point of view, this corresponds to the  introduction of 
orientifold planes. 
Tadpole cancellation then typically requires the introduction of an open sector,
corresponding to D-branes. 
In ten dimensions, there are two ways of obtaining an anomaly-free
type-I string from an orientifold projection of type-IIB.
The supersymmetric case corresponds to the introduction of an $O_-$-plane
(with negative tension and negative charge) and 
32 D9-branes, with a resulting gauge group  $SO(32)$ \cite{gs}. 
The cancellation
of the overall tension and charge of the configuration corresponds to the 
cancellation of dilaton and RR tadpoles. The resulting theory 
is ${\cal N}=1$ supersymmetric, and the massless spectrum contains the 
gravity multiplet from the closed sector and an $SO(32)$ Yang-Mills multiplet
from the open sector.
The second possibility corresponds to a change of sign of the tension
and the charge of the orientifold plane, so that RR tadpole cancellation 
requires the addition of 32 anti-D9 branes, with a resulting gauge 
group $USp(32)$ \cite{sugimoto}. 
The overall tension of the configuration does not vanish, so that the resulting 
theory has a dilaton tadpole. Nevertheless, the theory is anomaly-free, as a 
consequence of the vanishing of the RR tadpole \cite{pc,bm}. The 
spectrum is not supersymmetric \cite{bsb}, and more precisely the closed sector is not modified, 
still describing at the massless level the ${\cal N}=1$ gravity multiplet,
while the massless fermions in the open sector are not in the 
adjoint but in the antisymmetric representation of $USp(32)$. 
The gravitino couplings can then only be consistent if supersymmetry is non-linearly realized in
the open sector. Since the 
antisymmetric representation of symplectic groups is reducible, the massless spectrum 
contains a spinor that is an $USp(32)$ singlet, and this spinor is the goldstino of 
the non-linearly realized supersymmetry \cite{dm}.
 
In \cite{dm} the low-energy effective action for the $USp(32)$ model was 
built requiring that supersymmetry is realized in the closed sector
and spontaneously broken in the open sector. The couplings of the bulk (supersymmetric) 
fields to the goldstino were shown to have a geometric origin. More precisely,
adding to the bulk fields suitable terms containing the goldstino, the supersymmetry
transformation of these composite fields turns out to have the form of a general
coordinate transformation, plus additional gauge transformations in the case 
of gauge fields. The only exception are some 
Wess-Zumino-like terms resulting from the supersymmetrization of the 
Chern-Simons couplings. This is due to the fact that the field strength of the 
RR 2-form contains the Chern-Simons 3-forms coming from the vectors, so that 
it contains both bulk and brane fields. 
In \cite{pr} it was then shown that also these couplings 
have a geometric interpretation, provided one moves to a dual formulation, in which the RR
2-form is replaced with its dual 6-form, and the Chern-Simons term is replaced 
by a Wess-Zumino term $B \wedge F^2$.

The open sector of the effective action for type-I strings corresponds
to the first order in the low-energy expansion of the D9-brane action in a type-I
background. In general D-brane actions in type-II backgrounds consist of 
two terms: the first is the Dirac-Born-Infeld term, describing the coupling of the
brane to gravity and to the NS 2-form, while 
the second is the Wess-Zumino term, describing the coupling 
of the brane to the RR forms \cite{borninfeld}. The method for constructing actions for 
super D-branes is known in the literature \cite{dbranes,dbranes2,bt,aps1,aps2,bdail}. A basic ingredient for this 
construction is $\kappa$-symmetry, and after $\kappa$-gauge fixing one can show that
only half of the supersymmetries are linearly realized \cite{aps2}.
We are interested in this letter in the supersymmetric
brane action for a D9-brane of type IIB. We will comment about 
the relation between the general method of constructing this action and 
the method of \cite{dm,pr} of coupling non-supersymmetric matter in a 
supersymmetric background.
The action for a D9-brane in type-I string theory is obtained performing a 
truncation of the D9-brane action of type-IIB string theory \cite{truncation}. 
The remarkable result is that there are two possibilities of performing this 
truncation. In the flat background, with all bulk fields put to zero,
one case reduces to the 
Volkov-Akulov action \cite{va}, and the other case to a constant. 

In this letter we extend the results of \cite{truncation} to a generic background. 
We find that the couplings of the brane to the bulk have a geometric formulation. 
Also in the curved case there are two possibilities of performing the truncation.
In one case we get a dilaton tadpole and a RR tadpole plus 
goldstino couplings, while in the other case the goldstino couplings vanish and we are left with
a dilaton and a RR tadpole. We interpret this result as the equivalent of the 
string result: the two different truncations correspond to the two different choices 
of the relative sign of tension and charge of orientifold plane and D9-branes. 
The first case corresponds to the non-supersymmetric one, in which
the orientifold plane and the D-brane have both positive tension, and in the case of 32 coincident 
D9-branes it should give rise to the low-energy action \cite{dm,pr} of the $USp(32)$ 
model. The second case, in which the goldstino disappears, corresponds to the 
case in which the orientifold plane has negative tension, and in the case of 32
coincident D9-branes it should give rise to the low-energy action of the supersymmetric
$SO(32)$ superstring.
We are considering here only a single D-brane, while a more precise analogy between 
this low-energy analysis and the string result, requiring an extention to the non-abelian case
\cite{nonabelian}, is currently under investigation.

The letter is organized as follows. In Section 2 we review the results of \cite{truncation} 
in order to write the D9-brane action in a generic IIB background.
In Section 3 we perform the type-I truncation on the D9-brane action in a generic 
truncated background. Section 4 contains the conclusions.

\section{The D9-brane action}
Following the notations of \cite{truncation}, 
the supersymmetry transformations of the IIB bulk fields are
\bea
& & \d e_\m{}^a = \bar{\e} \G^a \psi_\m \quad , \nonumber \\
& & \d \psi_\m = D_\m \e -\frac{1}{8} H_{\m\n\r} \G^{\n\r} \sigma^3 \e 
+ \frac{1}{16} e^\phi \sum_{n=1}^6 \frac{1}{(2n-1)!} G_{\m_1 ... \m_{2n-1}}^{(2n-1)} 
\G^{\m_1 ... \m_{2n-1}}
\G_\m {\cal P}_n \e \quad , \nonumber \\
& & \d B_{\m\n} = 2 \bar{\e } \sigma^3 \G_{[\m} \psi_{\n ]} \quad , \nonumber \\
& & \d B^{(10)}_{\m_1 ...\m_{10}} = e^{-2 \phi} \bar{\e} \sigma^3 ( 10 
\G_{[ \m_1 ...\m_9 } \psi_{\m_{10}]} - \G_{\m_1 ... \m_{10}}  \l ) \quad , \nonumber \\
& & \d C^{(2n-2)}_{\m_1 ...\m_{2n-2}} =- (2n-2) e^{- \phi} \bar{\e} {\cal P}_n 
\G_{ [ \m_1 ... \m_{2n-3}}  (\psi_{\m_{2n-2}] }- \frac{1}{2(2n-2)} \G_{\m_{2n-2}]} \l )\nonumber \\
& & \qquad \qquad \quad +\frac{1}{2} (2n-2)(2n-3) C^{(2n-4)}_{[ \m_1 ... \m_{2n-4}}
\d B_{\m_{2n-3} \m_{2n-2}]} \quad , \nonumber \\ 
& & \d \l = \de_\m \phi \G^\m \e -\frac{1}{12} H_{\m\n\r} \sigma^3 \G^{\m\n\r} \e 
+\frac{1}{4} e^\phi \sum_{n=1}^6 \frac{n-3}{(2n-1)!} \G^{(2n-1)}_{\m_1 ...\m_{2n-1}}
{\cal P}_n \G^{\m_1 ... \m_{2n-1}} \e \quad , \nonumber \\
& & \d \phi = \frac{1}{2} \bar{\e} \l  \quad ,\label{susytransf}
\eea 
where ${\cal P}_n$ is $\sigma^1$ for $n$ even and $i \sigma^2$ for $n$ odd.
We are neglecting terms cubic in the fermions in the case of the transformations of the spinors.
We have introduced the field strengths for
the RR fields and their duals, related by duality according to the relations
\be
G^{(7)} = - * G^{(3)} \quad , \qquad G^{(9)} =  * G^{(1)} \quad , \qquad G^{(5)} =  * G^{(5)} 
\quad .
\ee
Moreover, the field strengths are defined through the relations
\bea
& & H = dB \nonumber \\
& & G^{(2n+1) } = dC^{(2n)} - H C^{(2n-2)} \quad , 
\eea
and the gauge transformations of the fields are
\bea
&  & \d B = d \L_{NS} \quad ,\nonumber \\
& & \d B^{(10)} = d \L_{NS}^{(10)} \quad ,\nonumber \\
& & \d C^{(2n)} = d \L_{RR}^{(2n-1)} - \L_{RR}^{(2n-3)} H \quad ,
\eea
so that the field strengths are gauge invariant.

We now want to introduce the action for the super D9-brane in curved IIB superspace.
Following \cite{bt}, we introduce the worldvolume fields as the supercoordinates 
\be
Z^M (\xi^i ) = ( x^\m (\xi^i ) , \theta^{{\a} I }(\xi^i ) ) 
\ee
defining the position of the brane in superspace. Here $\xi^i$ are the worldvolume 
coordinates, and $\m=0,...,9$, $i=0,...,9$, $\a=1,...,32$ and $I=1,2$. The spinors
$\th^I$ are both left-handed. 
We denote with $V^i (\xi )$ the abelian worldvolume vector.
The bulk superfields are denoted with
\be
\lbrace \phi , E_M{}^A , B_{MN} , B_{M_1 ...M_{10}}, C^{(2n-2)}_{M_1 ...M_{2n-2}} \rbrace
\quad , \qquad n= 1,...,6 \quad ,
\ee
and the action is 
\be
S = S_{DBI} + S_{WZ} = - \int_{M_{10}} d^{10} \xi e^{-\phi} \sqrt{ - \det (g +{\cal F })}
+ \int_{M_{10}} C e^{\cal F} \quad ,
\ee
where 
\be
{\cal F}_{ij } = F_{ij} + B_{ij } \quad , \qquad g_{ij} = E_i{}^a E_j{}^b \eta_{ab} \quad , \qquad 
B_{ij} = \de_i Z^M \de_j Z^N B_{MN} \quad .
\ee
and 
\be
C = \sum_{n=1}^6 (-1)^{n+1} C^{(2n-2)} \quad , \qquad C^{(2n-2)}= \frac{1}{(2n-2)!} 
d Z^{M_1} ... d Z^{M_{2n-2}}  C^{(2n-2)}_{M_1 ... M_{2n-2}} \quad .
\ee

where $\g_i = \G_\m \de_i X^\m$. 
The action is also $\kappa$-symmetric, and one can choose the  $\kappa$-gauge by
setting $\theta_2 = 0$, so that the resulting brane fields are a single Majorana-Weyl 
spinor and a vector, and the resulting spectrum is ${\cal N} =1 $ supersymmetric.   
In a flat background, 
the pullback of the metric reads
\be
g_{ij} = \eta_{ij} + \bar{\theta} \g_{(i } \de_{j )}  \theta \quad , \qquad \eta_{ij} \equiv \de_i x^\m 
\de_j x_\m \quad ,
\ee
while the NS 2-form becomes
\be
B_{ij} =  \bar{\theta} \sigma^3 \g_{[i} \de_{j]} \theta
\ee
up to higher order in the fermions. The RR fields become 
\be
C^{(2n+2)}_{i_1 ...i_{2n+2}} = - (n+1)  e^{- \phi} \bar{\theta} {\cal P}_n
\g_{[ i_1 .... i_{2n+1}} \de_{i_{2n+2}]}\theta \quad ,
\ee 
so that the resulting 
lagrangian, 
up to quartic terms in the fermions, is \cite{nonabelian}
\bea
& & {\cal L} = - \sqrt{- \det \eta} ( 1+ \frac{1}{2} \bar{\theta} \g^i \de_i \theta + 
\frac{1}{2} \bar{\theta} \sigma^3 \g_{i} \de_\j \theta F^{ij} \nonumber \\
& & \quad \qquad + \frac{1}{4} F^{ij} F_{ij} + \frac{1}{2} \bar{\theta} \g_i \de_j 
T^{ij} +...)\nonumber \\
& & \quad \qquad + e^{i_1 ... i_{10}} \sum_{k=0}^4
\frac{(-1)^k }{2^{k+1} k! (9-2k)! } \bar{\theta}{\cal P}_k \g_{i_1 ... i_{9-2k}} \de_{i_{10-2k}}
\theta ({\cal F}^k)_{i_{11-2k} ... i_{10}} \quad ,
\label{d9branelag}
\eea
where $T^{ij}$ is the energy-momentum tensor
\be
T^{ij} = F^{ik} F_k{}^j +\frac{1}{4} \eta^{ij} F_{kl} F^{kl} \quad .
\ee
The (global) supersymmetry transformations of the brane fields are
\bea
& & \d x^\m = -\frac{1}{2} \bar{\e} \G^\m \theta \quad , \nonumber \\
& & \d \theta = -\e \quad , \nonumber \\
& & \d V_i = -\frac{1}{2} \bar{\e} \g_i \s^3 \theta -\frac{1}{2}\bar{\e } \g^j \theta  F_{ji} \quad . 
\eea
These transformations act on  $g_{ij}$ as a general coordinate transformation of parameter
\be
a_i = -\frac{1}{2}\bar{\e } \g_i \theta \quad ,\label{gct}
\ee
and on $B_{ij}$ as a tensor gauge transformation of parameter\footnote{The general coordinate
transformation on $B$ contributes to terms quartic in the fermions and so it is not visible at this
order.}
\be
\L_i = \frac{1}{2} \bar{\e} \sigma^3 \g_i \theta \quad .\label{tgt}
\ee
The variation of the RR fields is also a tensor gauge transformation. 
From the supersymmetry transformation of $V_i$ one then deduces that ${\cal F}_{ij}$ is 
invariant under supersymmetry.  
The term $a^j F_{ji}$ in the transformation of $V_i$ 
is a general coordinate transformation of parameter $a$ plus a gauge transformation
of parameter 
\be
\L = - a^i V_i \quad .
\ee 

There is a natural geometrical way of understanding why the lagrangian (\ref{d9branelag})
is supersymmetric. 
Let us briefly review how a single spinor can be treated {\it \`a la} 
Volkov-Akulov \cite{va} as a goldstino of global supersymmetry. Let us 
restrict our attention to the ten dimensional case, considering a 
Majorana-Weyl fermion $\th$ with the 
supersymmetry transformation
\be
\d \th =- \e - \frac{1}{2}(\bar{\e} \g^\m \th ) \de_\m \th \quad . \label{va} 
\ee
The commutator of two such transformations is a translation,
\be
[\d_1 , \d_2 ] \th = (\bar{\e}_2 \g^\m \e_1 )  \de_\m \th \quad ,
\ee
and thus eq. (\ref{va}) provides a realization of supersymmetry.
The 1-form
\be
e_\m{}^a= \d_\m^a +\frac{1}{2}(\bar{\th} \g^a \de_\m \th )\quad ,
\ee
transforms under supersymmetry as 
\be
\d e^a = L_a e^a \quad ,
\ee
with $L_a$ the Lie derivative with respect to
\be
a_\m= -\frac{1}{2}(\bar{\e}\g_\m \th ) \quad . 
\ee
The action of supersymmetry on $e$ is thus a general coordinate 
transformation, with a parameter depending on $\th$, and therefore
\be
{\cal L} = -\det e 
\ee
is clearly an invariant Lagrangian. 
Using the same technique, for a generic field $A$ that transforms 
under supersymmetry as 
\be
\d A = L_a A \quad , \label{susygct}
\ee
defining the induced metric as $g_{\m\n}=e_\m{}^m e_{\n m}$, a 
supersymmetric lagrangian in flat space is determined by
the substitution
\be
{\cal L}(\eta,A) \rightarrow e{\cal L}(g,A) \quad .
\ee
Our lagrangian (\ref{d9branelag}) generalizes this construction 
in order to take into account additional gauge symmetries. 

It is then natural to generalize this construction to a generic 
background. One must construct from the bulk fields quantities whose 
supersymmetry variations are general coordinate transformations with 
the parameter $a$ plus additional gauge transformations. Supersymmetry
guarantees that this way of constructing the D9-brane action coincides
with the superspace construction of \cite{bt}\footnote{See \cite{bdail} for a similar construction 
in the case of a generic p-brane.}.
We thus define 
\bea
& & \hat{\phi} = \phi +\frac{1}{2} \bar{\theta} \l - \frac{1}{48} 
H_{ijk} \bar{\theta} \g^{ijk} \sigma^3 \theta \nonumber \\
& & \quad \quad +\frac{1}{16} e^\phi \sum_{n=1}^6 \frac{n-3}{(2n-1)!} 
G^{(2n-1)}_{i_1 ... i_{2n-1}} \bar{\theta} \g^{i_1 ... i_{2n-1}}{\cal P}_n \theta \quad ,
\eea
whose supersymmetry transformation is a general coordinate transformation 
with the correct parameter $a_i$ given in (\ref{gct}), up to higher order fermi terms.
With the same technique, one can construct all the other hatted fields \cite{dm,pr},
so that the resulting D9-brane action in a generic type-IIB background is 
\be
S = S_{DBI} + S_{WZ} = - \int_{M_{10}} d^{10} \xi e^{-\hat{\phi}} \sqrt{ - \det (\hat{g} +{\cal F })}
+ \int_{M_{10}} \hat{C} e^{\cal F} \quad ,
\ee
where
\be
{\cal F}_{ij } = F_{ij} + \hat{B}_{ij }\quad .
\ee
We will discuss in more detail this construction in the next section, in which we 
truncate the D9-brane action to a type-I background.

\section{Type-I truncations of the D9-brane action} 
The type-I truncation of type-IIB supergravity is achieved imposing \cite{truncation}
\bea
& & C^{(2n-2)} =0 \quad , \qquad n=1,3,5 \quad , \nonumber \\
& & B =0 \quad ,\nonumber \\
& & B^{(10)} =0 \quad ,\nonumber \\
& & (1\pm \sigma^1 ) f =0 \quad ,
\eea 
where we have denoted with $f$ the gravitino and the dilatino. 
The surviving bosonic fields are thus the dilaton, the metric, the RR 2-form and its dual,
and the RR 10-form, while the two different signs in the projection of the fermions 
indicate that there are two possible type-I truncations. 
The supersymmetry transformation rules are obtained truncating eq. (\ref{susytransf}), and turn
out to be
\bea
& & \d e_\m{}^a = \bar{\e} \G^a \psi_\m \quad , \nonumber \\
& & \d \psi_\m = D_\m \e \mp \frac{1}{8 \cdot 3!} e^\phi G^{(3)}_{\n\r\s} \G^{\n\r\s}\G_\m \e \quad ,
\nonumber \\ 
& & \d C^{(2)}_{\m\n} = \pm 2 e^{- \phi} \bar{\e} \G_{[\m} (\psi_{\n]} -\frac{1}{4}\G_{\n]}\l )\quad ,
\nonumber \\
& & \d C^{(10)}_{\m_1 ... \m_{10}} = \pm 10 e^{-\phi } \bar{\e} \G_{[ \m_1 ...\m_9} (
\psi_{\m_{10} ]} - \frac{1}{20} \G_{\m_{10}]} \l ) \quad , \nonumber \\
& & \d \l = \de_\m \phi \G^\m \e \pm \frac{1}{2 \cdot 3!} e^\phi G_{\m\n\r} \G^{\m\n\r} \e
\quad , \nonumber \\
& & \d \phi = \frac{1}{2} \bar{\e} \l  \quad .\label{susytransftrunc}
\eea 
Following \cite{truncation}, we then perform the same truncation on the D9-brane action.
The brane fields are projected according to
\bea
& & V_i =0 \quad , \nonumber \\
& & (1 \pm \sigma^1 ) \theta =0 \quad .
\eea
The lower sign choice leads to no surviving $\kappa$-symmetry.

We first perform the truncation of the D9-brane action in a flat background. 
Since the truncation puts both the vector $V_i$ and the NS 2-form $B$ to zero, 
all terms containing ${\cal F}$ in the brane action vanish. We are thus left 
with $g_{ij}$ and $C^{(10)}$, and we have
\bea
& & g_{ij} = \eta_{ij} + \bar{\theta} \g_{(i} \de_{j)} \theta \quad , \nonumber \\
& & C^{(10)}_{i_1 ... i_{10}} = \pm 5 \bar{\theta} \g_{[ i_1 ... i_9} \de_{i_{10}]} \theta \quad .
\eea
The lagrangian becomes
\be
{\cal L} = - \sqrt{- \det \eta} ( 1 + \frac{1}{2} \bar{\th} \g^i \de_i \th )
\mp \frac{5}{10!} \e^{i_1 ... i_{10}} \bar{\theta} \g_{[ i_1 ... i_9} \de_{i_{10}]} \theta \quad .
\label{trunclag}
\ee
In \cite{truncation} it was observed that using the properties of the ten-dimensional $\g$ matrices
and the fact that left-handed spinors (like $\theta$ and $\psi$) satisfy the relation
\be
\g_{i_1 ... i_{10}} f = - \e_{i_1 ... i_{10}} f \quad ,
\ee
the last term in (\ref{trunclag}) is equal to 
\be
\pm \frac{1}{2} \sqrt{ -\det \eta} \bar{\th} \g^i \de_i \th \quad .
\ee
The two choices of sign give then two different results: the upper sign gives just 
\be
- \sqrt{ -\det \eta}\quad ,
\ee
so that the $\th$ dependence cancels, while the lower sign gives
\be
-\sqrt{ -\det \eta} ( 1 + \bar{\th} \g^i \de_i \th ) \quad ,
\ee
that is exactly the Volkov-Akulov lagrangian to lowest order in the fermi fields.

We now want to generalize this construction in the case of a generic type-I background.
In order to do this, we proceed like in the IIB case of the previous section. Starting 
from the supersymmetry transformations of the bulk fields of eq. (\ref{susytransftrunc}),
we define the hatted fields
\bea 
& & \hat{\phi} = \phi  +\frac{1}{2} \bar{\theta} \l \pm \frac{1}{8 \cdot 3!} e^{\phi }
G^{(3)}_{ijk} \bar{\th} \g^{ijk} \th \quad ,\nonumber \\
& & \hat{g}_{ij} = g_{ij} + 2 \bar{\th} \g_{(i} \psi_{j)} + \bar{\th} \g_{(i} D_{j)} \th \nonumber \\
& & \quad \quad \mp \frac{1}{8} e^{\phi } G^{(3)}_{(i}{}^{kl} \bar{\th} \g_{j) kl} \th
\pm \frac{1}{16 \cdot 3} e^{\phi} G^{(3)}_{klm} \bar{\th} \g^{klm} \th g_{ij} \quad , \nonumber \\
& & \hat{C}^{(10)}_{i_1 ... i_{10}} = C^{(10)}_{i_1 ... i_{10}} \pm 10
e^{-\phi} \bar{\th} \g_{[i_1 ... i_9} \psi_{i_{10}]} \mp \frac{1}{2} e^{-\phi} \bar{\th}
\g_{i_1 ... i_{10}} \l \nonumber \\
& & \quad \quad \pm 5 e^{-\phi } \bar{\th} \g_{[ i_1 ... i_9} D_{i_{10}]} \theta
- 15 \bar{\th} \g_{[ i_1 ... i_7} \th G_{i_8 i_9 i_{10}]} \quad ,
\eea
whose supersymmetry variations are general coordinate transformations of parameter $a$ plus 
an additional tensor gauge transformation of parameter
\be
\L_{i_1 ...i_9}= - a^j C^{(10)}_{j i_1 ... i_9} \mp e^{-\phi} \bar{\th} \G_{i_1 ... i_9} \e 
\ee
in the case of $\hat{C}^{(10)}$. 
Substituting these fields in the truncated brane lagrangian
\be
{\cal L} = -  e^{-\hat{\phi}} \sqrt{ - \det \hat{g} }
- \hat{C}^{(10)} \quad ,
\ee
the upper sign choice gives
\be
-e^{- \phi} \sqrt{- \det g}  - C^{(10)} \quad ,
\label{so32}
\ee
while the lower sign choice gives
\be
-e^{-\phi} \sqrt{ -\det g} (1 +2 \bar{\th} \g^i \psi_i + \bar{\th} \g^i D_i \th
- \bar{\th} \l -\frac{1}{24} e^\phi G^{(3)}_{ijk} \bar{\th} \g^{ijk} \th ) -  C^{(10)}\quad .
\label{usp32}
\ee 
All these results are up to quartic terms in the fermions and up to higher order terms in derivatives.

The upper sign choice, then, corresponds to the vanishing of all the terms containing 
the goldstino, so that the resulting spectrum is supersymmetric. 
The brane action (\ref{so32}) contains a dilaton tadpole and a RR tadpole. It is then natural to
argue that this projection corresponds to the case in which the relative sign between tension
and RR charge of the brane is the same as the relative sign between tension and RR charge of the
orientifold plane. Adding 32 D9-branes and performing a non-abelain generalization of 
this projection, the overall tension
and charge should then cancel, giving rise to the low-energy action for the $SO(32)$ type-I string.
The projection, in the non-abelian case, then, should preserve the antisymmetric representation
for both the vector $V$ and the spinor $\th$.
The lower sign choice corresponds to the curved generalization of the Volkov-Akulov action. 
The resulting spectrum is non-supersymmetric in the brane sector, or more precisely
supersymmetry is non-linearly realized on the brane. The brane action (\ref{usp32}) contains 
a dilaton tadpole and a RR tadpole, but in this case the projection corresponds
to the case in which the relative sign between tension
and RR charge of the brane is opposite to the  relative sign between tension and RR charge of the
orientifold plane. This means that no configuration of branes can cancel both tadpoles. The inclusion
of 32 (anti)D9-branes cancels the overall charge, while the dilaton tadpole remains, and this corresponds
to the low-energy action of the $USp(32)$ type-I string. In this case 
the projection, in the non-abelian case, should preserve the symmetric representation
for the vector $V$ and the antisymmetric one for the spinor $\th$.

\section{Conclusions}
The aim of this letter was to clarify the relation between the type-I truncation of type-IIB 
supergravity when D9-branes are present and the low-energy effective action of ten-dimensional
type-I strings. It was shown in \cite{truncation} that there 
are two possible type-I truncations, that give two different results in a flat background, namely in one 
case one gets just a constant, and the dependence on $\th$ vanishes, 
while in the other case one gets a Volkov-Akulov action for $\th$. The generalization 
of these results to any background shows that these two truncations correspond to the two possible 
choices of sign of the tension of the orientifold plane with respect to its charge. 
In the first case the relative sign between tension
and RR charge of the brane is the same as the relative sign between tension and RR charge of the
orientifold plane, so that the spectrum is supersymmetric and the brane action reduces to a dilaton and 
a RR tadpole.
In the second case  the relative sign between tension
and RR charge of the brane is opposite to the  relative sign between tension and RR charge of the
orientifold plane. The spectrum is non-supersymmetric on the brane, and $\th$ is the goldstino of 
the non-linearly realized local supersymmetry. 
In the case of 32 
9-branes, then, these two results must correspond to the two anomaly-free and tachyon-free type-I 
string theories in ten-dimensions, namely the $SO(32)$ and the $USp(32)$ strings.
Non-abelian generalizations of brane actions were costructed in \cite{nonabelian}. It is interesting 
to generalize these truncations to the non-abelian case. The Wess-Zumino term of the 
truncated non-abelian action contains the RR 2-form $C^{(2)}$ and its dual 6-form $C^{(6)}$, together
with the 10-form $C^{(10)}$, so that we expect that, in analogy with \cite{pr}, there can be 
subtleties related to the geometrization procedure in this case. 

In order to perform the truncation, two 10-forms were introduced in the fields of type-IIB supergravity.
Even though these fields have no dynamics, since they do not have a field strength, their presence is 
important for various reasons. In particular, the RR 10-form provides 
the RR charge for D9-branes, and so it is
fundamental in order to understand the consistency of type-I theories. S-duality implies 
the presence of the NS 10-form $B^{(10)}$ \cite{hull}, that provides the electric charge for 
NS9-branes \cite{ns9}.
It would then be interesting to study the S-dual of this truncation: this would correspond to a
heterotic truncation in the presence of NS9-branes. In \cite{truncation} the heterotic truncation 
of the D9-action was studied. In this case the vector is not projected out, and a single Majorana-Weyl
spinor survives in the brane sector. To the best of our knowledge, there is no string interpretation
of this result at the moment, and it is not clear to us how a brane configuration could give rise 
to an anomaly-free spectrum corresponding to this truncation.

\section*{Acknowledgements}
I am grateful to E. Bergshoeff for having pointed to my attention Ref. \cite{truncation} and for 
interesting discussions. This work is part of the research program of the Foundation for
Fundamental Research on Matter (FOM) and the Netherlands Organization 
for Scientific Research (NWO).

\end{document}